\documentclass[twocolumn,showpacs,preprintnumbers,amsmath,amssymb]{revtex4}

\usepackage{graphicx}
\usepackage{dcolumn}
\usepackage{bm}

\begin{document}

\preprint{APS/123-QED}

\title{$S$ =1/2 ferromagnetic Heisenberg chain in a verdazyl-based complex}

\author{N. Uemoto$^1$, Y. Kono$^2$, S. Kittaka$^2$, T. Sakakibara$^2$, T. Yajima$^2$, Y. Iwasaki$^1$, S. Miyamoto$^1$, Y. Hosokoshi$^1$, and H. Yamaguchi$^1$}
\affiliation{
$^1$Department of Physical Science, Osaka Prefecture University, Osaka 599-8531, Japan \\ 
$^2$Institute for Solid State Physics, the University of Tokyo, Chiba 277-8581, Japan\\}

Second institution and/or address\\
This line break forced

\date{\today}

\begin{abstract}
We present a model compound for the $S$=1/2 ferromagnetic Heisenberg chain composed of the verdazyl-based complex $[$Zn(hfac)$_2]$$[$4-Cl-$o$-Py-V-(4-F)$_2]$. 
$Ab$ $initio$ MO calculations indicate a predominant ferromagnetic interaction forming an $S$=1/2 ferromagnetic chain. 
The magnetic susceptibility and specific heat indicate a phase transition to an AF order owing to the finite interchain couplings.
We explain the magnetic susceptibility and magnetization curve above the phase transition temperature based on the $S$=1/2 ferromagnetic Heisenberg chain.
The magnetization curve in the ordered phase is described by a conventional AF two-sublattice model.
Furthermore, the obtained magnetic specific heat reproduces the almost temperature-independent behavior of the $S$=1/2 ferromagnetic Heisenberg chain.
In the low-temperature region, the magnetic specific heat exhibits $\sqrt{T}$ dependence, which is attributed to the energy dispersion in the ferromagnetic chain.
\end{abstract}

                                           
\maketitle
\section{INTRODUCTION}
One-dimensional (1D) spin systems have been intensively investigated  both experimentally and theoretically, over the last several decades.
In particular, the $S$ = 1/2 antiferromagnetic (AF) Heisenberg 
chain is of fundamental importance in strongly correlated quantum many-body systems, because it is one of the few systems for which a nontrivial ground state is precisely known. 
The ground state of the $S$ = 1/2 AF Heisenberg chain is described by a Tomonaga-Luttinger liquid, which is a quantum critical state with a $k$-linear energy dispersion~\cite{1DAF}.
Such linear dispersion results in the well-known $T$-linear behavior of the low-temperature specific heat.
Although many examples of the $S$ = 1/2 AF Heisenberg chain have been reported so far, the number of examples of $S$ = 1/2 ferromagnetic Heisenberg chain is much smaller.
Nevertheless, theoretical investigation of the $S$ = 1/2 ferromagnetic Heisenberg chain have been performed over many decades.
The ground state is a ferromagnetic order, but there is no ordered state at any finite temperature.
Bethe demonstrated that the exact many-body wave function of such a system can be expressed analytically~\cite{bete}, and Bonner and Fisher studied the thermodynamic properties~\cite{BF}. 
Takahashi explained the low-temperature thermodynamic properties by using the modified spin-wave approximation~\cite{takahashi}.
It is confirmed that the magnetic susceptibility diverges as $T^{-2}$ with decreasing temperature and that the low-temperature specific heat exhibits a $\sqrt{T}$ dependence associated with the $k^2$ dependence of the energy dispersion~\cite{ferro1,ferro2,ferro3,ferro4}.  
In the case of an actual quasi-1D ferromagnetic system, theoretical study suggests that predominant 1D ferromagnetic interactions realize a dimensional crossover of Bose-Einstein condensation (BEC)~\cite{ferro_BEC}.

From the experimental point of view, several Cu complexes have been actively investigated as the best realizations of the $S$ = 1/2 ferromagnetic Heisenberg chain mainly in the 1980s~\cite{TMSO_PRB, DMSO_JAP, CHAB_PRB}.
In (C$_6$H$_{11}$NH$_3$)CuX$_3$ (X=Cl and Br), the almost temperature-independent specific heats are well explained by using ferromagnetic chain models, while low-temperature $\sqrt{T}$ dependence is not observed due to the appearance of large peaks associated with the phase transition to long-range-order (LRO)~\cite{CHAB_PRB}. 
Neutron scattering studies on these Cu complexes demonstrated the $k^2$ dependence of the spin-wave dispersion in the chain direction~\cite{DMSO_neut, CHAB_neut}. 
Although these Cu complexes have strong ferromagnetic chain interactions, their magnetic properties exhibit slightly anisotropic behavior.
Thus, organic radical compounds with isotropic nature owing to orbital quenching have attracted interest as a model system of ferromagnetic Heisenberg chain.
From the 1990s, several organic radical compounds, e.g. $\beta $-$p$-NPNN, $\gamma$-$p$-NPNN, BImNN, F4BImNN, and $\beta$-BBDTA-GaBr$_4$, have been reported to form the $S$ = 1/2 ferromagnetic Heisenberg chain~\cite{b-p-NPNN, r-p-NPNN, b-r-both, BImNN_CL, BImNN_uSR, F4BImNN_JACS, F4BImNN_uSR, BBDTA_PRB}.
Their magnetic properties in the ordered phase and vicinity of the phase transition temperature are examined in connection with the pursuit for organic magnet.

In order to design Heisenberg spin systems, we have also focused on organic radical compounds. 
One of the suitable candidates is verdazyl radical, which can exhibit a delocalized $\pi$-electron spin density in nonplanar molecular structures. 
The flexibility of the molecular orbitals in the verdazyl radical offers tunability of the intermolecular magnetic interactions by molecular design. 
Recently, we demonstrated that the verdazyl radical can form a variety of unconventional $S$ = 1/2 Heisenberg spin systems, such as
the quantum pentagon, random honeycomb, and fully-frustrated square lattice, which have not been realized in conventional inorganic materials~\cite{a26Cl2V, random, TCNQ_square}. 
Furthermore, in contrast to conventional organic radical systems, the verdazyl radical facilitates the formation of ferromagnetic interactions, and thus we have realized a variety of $S$ = 1/2 Heisenberg spin systems with 1D ferromagnetic interactions~\cite{3Cl4FV,2Cl6FV, PF6}.
For instance, antiferromagnetically coupled ferromagnetic chains form a two-dimensional honeycomb lattice in 2-Cl-6-F-V~\cite{2Cl6FV}, and the frustrated square lattice in ($o$-MePy-V)PF$_6$ contains  alternating ferromagnetic chains~\cite{PF6}.   
In 3-I-V, which forms a spin ladder with a predominant ferromagnetic-leg interaction, quasi-1D BEC was actually observed near the saturation field~\cite{3IV_kono}.

In this paper, we present a new verdazyl-based complex.
We successfully synthesized single crystals of $[$Zn(hfac)$_2]$$[$4-Cl-$o$-Py-V-(4-F)$_2]$  [hfac = 1,1,1,5,5,5-hexafluoroacetylacetonate, 4-Cl-$o$-Py-V-(4-F)$_2$ = 3-(4-Cl-2-pyridyl)-1,5-bis(4-fluorophenyl)-diphenylverdazyl].
$Ab$ $initio$ molecular orbital (MO) calculations indicate that a predominant ferromagnetic interaction forms an $S$=1/2 ferromagnetic chain. 
The magnetic susceptibility and specific heat show a phase transition to an AF order owing to the finite AF interchain couplings.
We explain the magnetic susceptibility, magnetization curve, and magnetic specific heat above the phase transition temperature based on the $S$=1/2 ferromagnetic Heisenberg chain.
Furthermore, we observe $\sqrt{T}$ dependence of the magnetic specific heat in the low-temperature region.

\section{EXPERIMENTAL AND NUMERICAL METHOD}
We synthesized 4-Cl-$o$-Py-V-(4-F)$_2$ through a conventional procedure for verdazyl radical~\cite{Kuhn}. 
A solution of $[$Zn(hfac)$_2]$$\cdot$2H$_2$O (510 mg, 0.99 mmol) in 15 ml of heptane was refluxed at 60$^\circ$C.
A solution of 4-Cl-$o$-Py-V-(4-F)$_2$ (380 mg, 0.99 mmol) in 2 ml of CH$_2$Cl$_2$ was slowly added, and stirring was continued for 1 h. 
After the mixed solution cooled to room temperature, a dark-green crystalline solid of $[$Zn(hfac)$_2]$$[$4-Cl-$o$-Py-V-(4-F)$_2]$ was separated by filtration and washed with heptane. The dark-green residue was recrystallized using acetonitrile at 10$^\circ$C.

Single crystal X-ray diffraction (XRD) experiment was performed by using a diffractometer with an imaging plate (R-AXIS RAPID, Rigaku) with graphite-monochromated Mo-K$\alpha$ radiation at room temperature. 
The single crystal XRD data are refined by using the SHELX software~\cite{Xray}. 
The structural refinement was carried out using anisotropic and isotropic thermal parameters for the nonhydrogen atoms and the hydrogen atoms, respectively. 
All the hydrogen atoms were placed at the calculated ideal positions

The magnetic susceptibility and magnetization curves were measured using a commercial SQUID magnetometer (MPMS-XL, Quantum Design) above 1.8 K and a capacitive Faraday magnetometer with a dilution refrigerator down to 80 mK.
The experimental results were corrected for the diamagnetic contribution of $-1.65{\times}10^{-4}$ emu mol$^{-1}$, which is determined based on the QMC analysis to be described and close to that calculated by Pascal’s method.
The specific heat was measured with a commercial calorimeter (PPMS, Quantum Design) using a thermal relaxation method above 1.9 K and a handmade apparatus by a standard adiabatic heat-pulse method with a dilution refrigerator down to about $\sim$70 mK. 
There is no significant difference in magnetic properties between the field directions owing to the isotropic $g$ value of $\sim$2.00 in verdazyl radical systems. 
Therefore, all experiments were performed using small randomly oriented single crystals.

$Ab$ $initio$ MO calculations were performed using the UB3LYP method with the basis set 6-31G in the Gaussian 09 program package.
The convergence criterion was set at 10$^{-8}$ hartree.
For the estimation of intermolecular magnetic interaction, we applied our evaluation scheme that have been studied previously~\cite{MOcal}. 

The QMC code is based on the directed loop algorithm in the stochastic series expansion representation~\cite{QMC2}. 
The calculations for the $S$ = 1/2 ferromagnetic Heisenberg chain was performed for $N$ = 1024 under the periodic boundary condition, where $N$ denotes the system size.
It was confirmed that there is no significant size-dependent effect.
All calculations were carried out using the ALPS application~\cite{ALPS,ALPS3}.

\section{RESULTS AND DISCUSSION}
\subsection{Crystal structure and magnetic model}
The crystallographic data for the synthesized $[$Zn(hfac)$_2]$$[$4-Cl-$o$-Py-V-(4-F)$_2]$ are summarized in Table I, and the molecular structure is shown in Fig. 1(a).
The verdazyl ring (which includes four N atoms), the upper two phenyl rings, and the bottom pyridine ring are labeled ${\rm{R}_{1}}$, ${\rm{R}_{2}}$, ${\rm{R}_{3}}$, and ${\rm{R}_{4}}$, respectively.
The dihedral angles of ${\rm{R}_{1}}$-${\rm{R}_{2}}$, ${\rm{R}_{1}}$-${\rm{R}_{3}}$, ${\rm{R}_{1}}$-${\rm{R}_{4}}$ are approximately 9$^{\circ}$, 51$^{\circ}$, and 9$^{\circ}$, respectively.
Each molecule has a delocalized $S$=1/2.
The result of the MO calculation for $[$Zn(hfac)$_2]$$[$4-Cl-$o$-Py-V-(4-F)$_2]$ molecule indicate that approximately 61 ${\%}$ of the total spin density is present on ${\rm{R}_{1}}$. 
Further, while ${\rm{R}_{2}}$ and ${\rm{R}_{3}}$ each account for approximately 14 ${\%}$ and 18 ${\%}$ of the relatively large total spin density, ${\rm{R}_{4}}$ accounts for less than 7 ${\%}$ of the total spin density.
Therefore, the intermolecular interactions are caused by the short contacts related to the ${\rm{R}_{1}}$, ${\rm{R}_{2}}$, and ${\rm{R}_{3}}$ rings.
Since Zn(hfac)$_2$ has a low spin density less than 1 ${\%}$ of the total spin density, it works as a spacer between verdazyl radicals, resulting in the low dimensionality of the spin model. 
We focus on the structural features related to the 4-Cl-$o$-Py-V-(4-F)$_2$ to consider intermolecular interactions.

We evaluated the intermolecular exchange interactions of all molecular pairs within 4.0 $\rm{\AA}$ through the $ab$ $initio$ MO calculations and found one predominant ferromagnetic interaction.
Its value was evaluated as $J/k_{\rm{B}}$ = $-10.7$ K, which is defined in the Heisenberg spin Hamiltonian given by $\mathcal {H} = J{\sum^{}_{<i,j>}}\textbf{{\textit S}}_{i}{\cdot}\textbf{{\textit S}}_{j}$, where $\sum_{<i,j>}$ denotes the sum over the neighboring spin pairs.
The molecular pair associated with this interaction is related by an $a$-glide reflection symmetry and has N-C short contact of 3.51 $\rm{\AA}$, as shown in Figs. 1(b) and (c).
Accordingly, a uniform $S$ = 1/2 ferromagnetic chain is formed by the expected interaction $J$ along the $a$-axis, as shown in Fig. 1(d).
The Zn(hfac)$_2$ acts as a spacer between the 1D chains, as shown in Fig. 1(e).
The other intermolecular interactions associated with interchain couplings are evaluated to be less than approximately 1/100 of $J$ in absolute values, which enhances the 1D character of the present spin model.
Considering strong dependence on the calculation method and basis set in the MO calculation, such small interchain couplings do not have enough reliability~\cite{MOseido}, and thus we evaluate the effective interchain couplings through a mean-field analysis of the experimental results, as discussed later.

\begin{table}
\caption{Crystallographic data for $[$Zn(hfac)$_2]$$[$4-Cl-$o$-Py-V-(4-F)$_2]$.}
\label{t1}
\begin{center}
\begin{tabular}{cc}
\hline
\hline 
Formula & C$_{29}$H$_{15}$ClF$_{14}$N$_{5}$O$_{4}$Zn\\
Crystal system & Orthorhombic \\
Space group & $Pbca$ \\
$a/\rm{\AA}$ & 10.7183(6) \\
$b/\rm{\AA}$ & 19.0465(10)  \\
$c/\rm{\AA}$ & 32.8749(16) \\
$V$/$\rm{\AA}^3$ & 6711.3(6) \\
$Z$ & 8 \\
$D_{\rm{calc}}$/g cm$^{-3}$ & 1.711 \\
Temperature & RT \\
Radiation & Mo K$\rm{\alpha}$ ($\lambda$ = 0.71075 $\rm{\AA}$) \\
Total reflections & 7625 \\
Reflection used & 3283 \\
Parameters refined & 487 \\
$R$ [$I>2\sigma(I)$] & 0.0673 \\
$R_w$ [$I>2\sigma(I)$] & 0.1697 \\
Goodness of fit & 1.002 \\
CCDC & 1889996 \\
\hline
\hline
\end{tabular}
\end{center}
\end{table}

\begin{figure*}[t]
\begin{center}
\includegraphics[width=36pc]{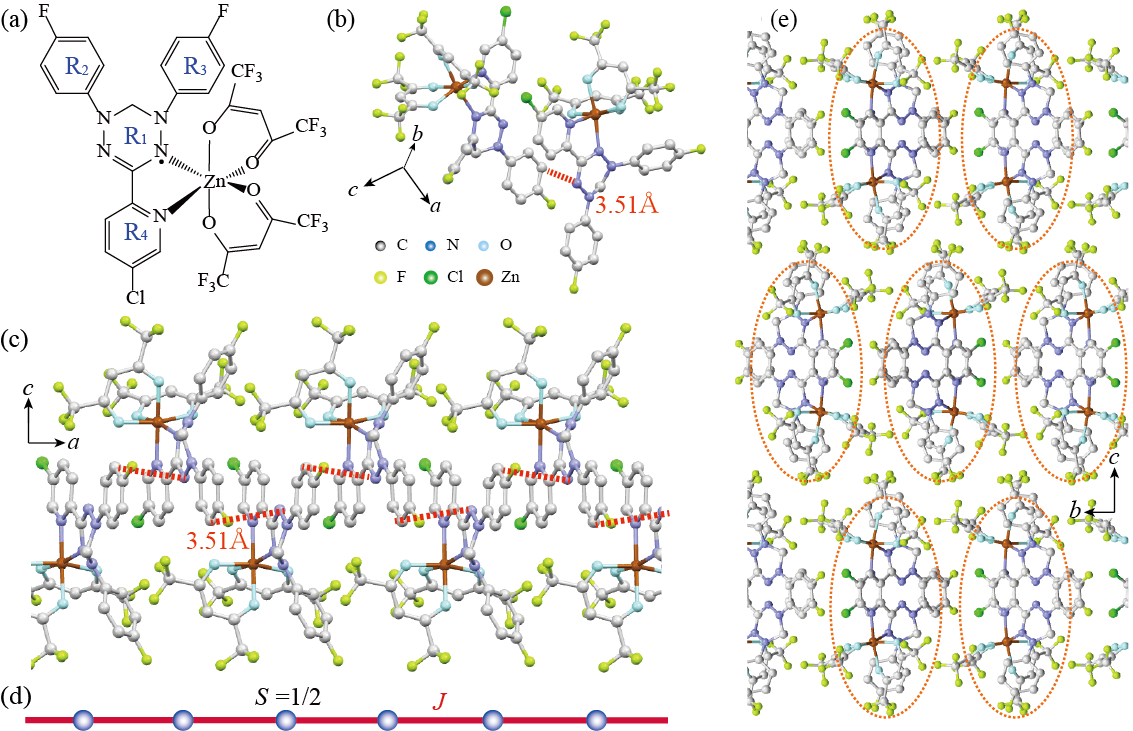}
\caption{(color online) (a) Molecular structure of $[$Zn(hfac)$_2]$$[$4-Cl-$o$-Py-V-(4-F)$_2]$. (b) Molecular pair associated with the dominant ferromagnetic interaction $J$. Hydrogen atoms are omitted for clarity. The broken line indicates N-C short contact. 
(c) Crystal structure forming a 1D chain along the $a$-axis, and (d) the corresponding $S$ = 1/2 ferromagnetic chain. 
(e) Crystal structure viewed parallel to the chain direction. The broken line encloses molecules comprising each chain structure.}\label{f1}
\end{center}
\end{figure*}

\begin{figure}[t]
\begin{center}
\includegraphics[width=18pc]{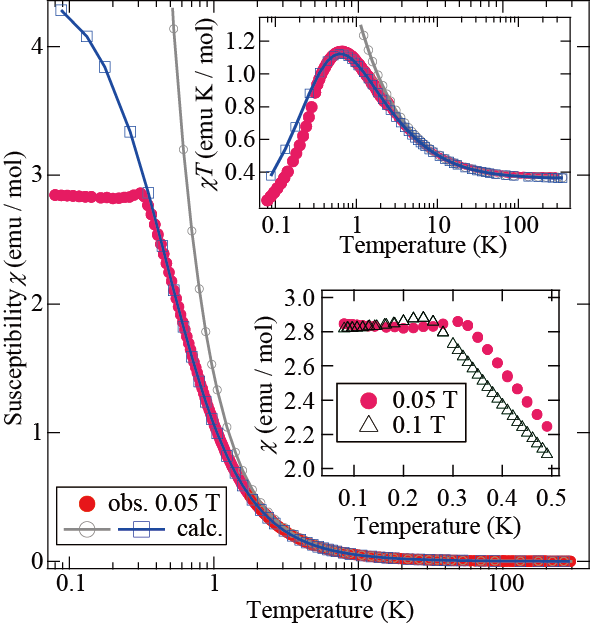}
\caption{(color online) Temperature dependence of magnetic susceptibility ($\chi=M/H$) of $[$Zn(hfac)$_2]$$[$4-Cl-$o$-Py-V-(4-F)$_2]$ at 0.05 T.
The upper inset shows the temperature dependence of $\chi T$.
The solid lines with open circles and open squares represent the calculated results for the isolated ferromagnetic chain and the ferromagnetic chain with the AF mean field, respectively. 
The lower inset shows $\chi$ in the vicinity of the phase transition temperature at 0.05 and 0.1 T.}\label{f3}
\end{center}
\end{figure}

\begin{figure}[t]
\begin{center}
\includegraphics[width=20pc]{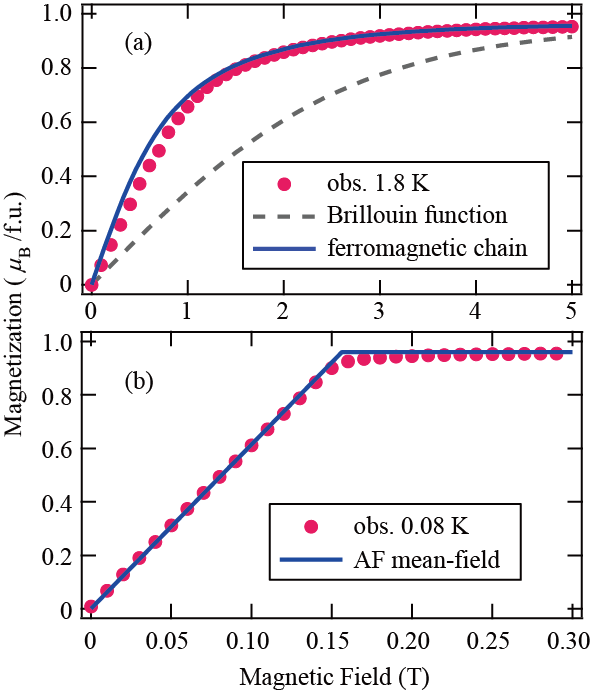}
\caption{(color online) Magnetization curve of $[$Zn(hfac)$_2]$$[$4-Cl-$o$-Py-V-(4-F)$_2]$ at (a) 1.8 K and (b) 0.08 K. The broken line represents the Brillouin function for $S$ = 1/2 at 1.8 K.
The solid lines represent the calculated results for (a) the $S$ =1/2 ferromagnetic chain at 1.8 K by using QMC method and for (b) the AF two-sublattice model at zero temperature by using the mean-field approximation.}\label{f3}
\end{center}
\end{figure}

\subsection{Magnetic susceptibility}　
Figure 2 shows the temperature dependence of the magnetic susceptibility ($\chi=M/H$) at 0.05 T. 
The contribution of $J_{\rm{F}}$ appears in the temperature dependence of $\chi$$T$, which increases with decreasing temperatures down to approximately 0.7 K, as shown in the upper inset of Fig. 2. 
At temperature above 2 K, the Curie-Weiss law is followed, $\chi$ = $C/(T-{\theta}_{\rm{W}})$.
The estimated Curie constant is about $C$ = 0.373 emu$\cdot$ K/mol, which is close to the expected value for noninteracting $S$ = 1/2 spins, and the Weiss temperature is estimated to be ${\theta}_{\rm{W}}$ = 3.2 K. 
Below 0.7 K, $\chi$$T$ decreases with decreasing temperature, indicating the existence of additional weak AF interactions.
Furthermore, we observe a discontinuous change in $\chi$ at 0.31 K, which indicates a phase transition to a three-dimensional LRO.
In the ordered phase, $\chi$ becomes almost temperature independent, which suggests that a spin-flop transition field is lower than 0.05 T as will be discussed later.
As shown in the lower inset of Fig. 2, the discontinuous change shifts to a lower temperature at a higher field of 0.1 T, which indicates that the ordered state is an AF ordering of the ferromagnetic chains.  

The MO calculations show the formation of the $S$ = 1/2 ferromagnetic chain, and the one-dimensionality is enhanced owing to the almost nonmagnetic Zn(hfac)$_2$ between the chain structures.
Accordingly, we calculated the magnetic susceptibility ${\chi}_{\rm{QMC}}$ for the $S$ = 1/2 ferromagnetic Heisenberg chain by using the QMC method.
Given the radical-based compound, we assume the isotropic $g$-value of 2.00.
The experimental result exhibits the contribution of weak AF interchain interactions as represented by the broad peak of $\chi$$T$, and thus we consider an AF mean field to reproduce the observed behavior.
The mean-field approximation is represented as:
\begin{equation}
\mathcal {\chi} = \frac{{\chi}_{\rm{QMC}}}{1+(zJ'/Ng^{2}{\mu}_{B}^{2}){\chi}_{\rm{QMC}}} 
\end{equation}
where $z$ is the number of nearest-neighbor spins, $J'$ is the interchain exchange interaction, $N$ is the number of spins, and ${\mu}$$_B$ is the Bohr magneton. 
We obtained good agreement between the experimental and calculated resluts above the phase transition temperature, including the broad peak of $\chi$$T$, by using the parameters $J/k_{\rm{B}}$ = $-$8.8 K and $zJ'/k_{\rm{B}}$ = 0.32 K, as shown in Fig. 2 and its upper inset.
A clear AF contribution can be confirmed by comparing the calculated results of the isolated ferromagnetic chain and the ferromagnetic chain with the AF mean field, as shown in Fig. 2 and its upper inset.
The Weiss temperature for the 1D chain is given by ${\theta}_{\rm{W}}$ = $-2JS(S+1)/3k_{\rm{B}}$ from the mean-field approximation, and we obtaine ${\theta}_{\rm{W}}$ = 4.4 K, which is close to the evaluation from the Curie-Weiss fitting.  

\subsection{Magnetization curve} 
Figure 3(a) shows the magnetization curve at 1.8 K above the phase transition temperature.
The saturation value of 0.96 $\mu_{\rm{B}}$/f.u. indicates that the purity of the radicals is approximately 96 $\%$.
The dominant ferromagnetic contribution can be confirmed by comparing the observed result and the Brillouin function for free $S$ = 1/2 spins, and the experimental result is well explained by the $S$ = 1/2 ferromagnetic Heisenberg chain with $J/k_{\rm{B}}$ = $-$8.8 K, as shown in Fig. 3(a).
The slight deviation is considered to originate from the contribution of weak AF interchain interactions as in the case of the magnetic susceptibility.

Figure 3(b) shows the magnetization curve at 0.08 K.
The samples used are the same as those for the magnetic susceptibility, and the experimental result corresponds to the magnetization in the AF ordered phase.
We observe an almost linear increase with increasing fields up to approximately 0.15 T.
The observed linear increase at 0.08 K indicates that the magnetic behavior in the ordered phase can be described by a classical AF two-sublattice model, in which the spins on each ferromagnetic chain form one sublattice moment.
The colinear two-sublattice is aligned along the easy-axis under zero-field conditions.
For the external field parallel to the the easy-axis, the discontinuous spin-flop phase transition occurs at a certain field. 
Above the spin-flop transition field, two sublattices are tilted to the field direction with equivalent angles. 
For the external field perpendicular to the easy-axis, two sublattices are tilted from the easy-axis with equivalent. 
In the present case, we have not observed anomalous behavior associated with the spin-flop transition in the magnetization curve of randomly oriented single crystals.
Thus, the transition field can be evaluated to be less than 0.01 T, which is consistent with the small magnetic anisotropy in the organic radical systems.
The magnetization curve becomes almost identical for any field direction above the spin-flop phase transition, yielding the almost temperature independent behavior of the magnetic susceptibility at 0.05 T in the ordered phase.
Accordingly, we can consider an isotropic spin system for the analysis even in the ordered phase. 
Two sublattices are coupled by the AF interchain interactions, and a mean-field approximation gives the magnetization curve at $T$ = 0, expressed as: 
$M_{\rm{mean}}=g^{2}{\mu _B}H/2zJ'$,
where $H$ is the external magnetic field. 
We obtained good agreement between the experimental and calculated results by using $zJ'/k_{\rm{B}}$ = $0.21$ K (as shown in Fig. 3(b)), which is close to the value obtained from the analysis of the magnetic susceptibility.

\begin{figure}[t]
\begin{center}
\includegraphics[width=18pc]{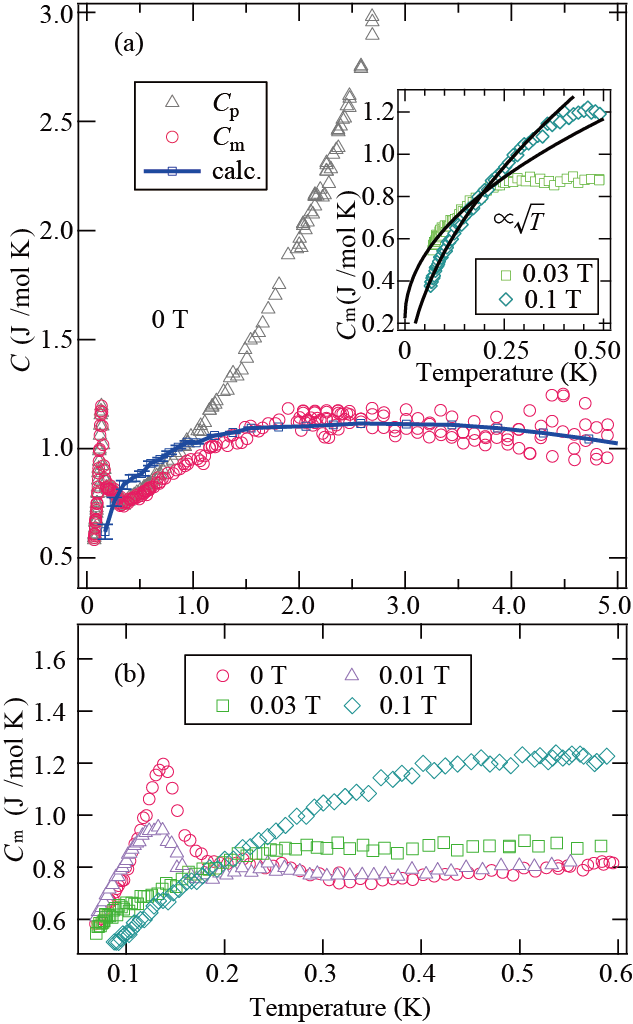}
\caption{(color online) Specific heat of $[$Zn(hfac)$_2]$$[$4-Cl-$o$-Py-V-(4-F)$_2]$. (a) Total specific heat $C_{\rm{p}}$ and its magnetic contribution $C_{\rm{m}}$ at 0 T. The solid line with squares represents the calculated result for the $S$=1/2 ferromagnetic Heisenberg chain. (b) Low-temperature region of  $C_{\rm{m}}$ at 0, 0.01, 0.03, and 0.1 T. The inset shows the $\sqrt{T}$ dependence of the low-temperature $C_{\rm{m}}$ at 0.03 and 0.1 T.}\label{f4}
\end{center}
\end{figure}

\subsection{Specific heat}
The experimental results for the specific heat $C_{\rm{p}}$ at zero-field clearly exhibit a $\lambda$-type sharp peak, which is associated with the phase transition to the AF LRO, as shown in Fig. 4(a).
Because we used different samples from those used for the magnetization measurements, the phase transition slightly shifts to a lower temperature.
We ascribe this disparity to a high sensitivity of the interchain couplings to the purity of the radicals and/or slight impurities. 
We evaluated the magnetic specific heat $C_{\rm{m}}$ by subtracting the lattice contribution $C_{\rm{l}}$ and assumed $C_{\rm{l}}$ in the low-temperature region approximated as: $C_{\rm{l}}=a_{1}T^{3}+a_{2}T^{5}+a_{3}T^{7}$.
The constants $a_{1}-a_{3}$ are determined to reproduce the following magnetic specific heat calculated by the QMC method.
As a result, $C_{\rm{l}}$ with the constants $a_{1}=0.12$, $a_{2}=-3.6\times10^{-3}$, and $a_{3}=5.7\times10^{-5}$ was evaluated.
The value of $a_{1}$ corresponds to a Debye temperature of 25 K, which is not very different from those for other verdazyl radical compounds, but is a slightly smaller value~\cite{3Cl4FV,2Cl6FV}.
This small Debye temperature reduces the applicable temperature region of the Debye's $T^3$ law, and thus we had to also consider $T^5$ and $T^7$ terms to evaluate the magnetic specific heat up to approximately 5 K.
The smaller value of Debye temperature is consistent with the molecular arrangement of the present compound, in which there is no large overlap of molecular orbitals causing strong AF coupling. 
We calculated the magnetic specific heat for the $S$ = 1/2 ferromagnetic Heisenberg chain with $J/k_{\rm{B}}$ = $-$8.8 K by using the QMC method and obtained good agreement between the experiment and calculation above the phase transition temperature, as shown in Fig. 4(a).
Here, we confirm that the calculated specific heat is consistent with those reported in previous theoretical works~\cite{BF,ferro2,ferro3,ferro4}.

The phase transition temperature decreases with increasing field, as shown in Fig. 4(b).
Such field dependence of the phase transition temperature is consistent with that for $\chi$ and also indicates the AF LRO.
At 0.03 and 0.1 T, the phase transition temperature becomes lower than the experimental temperature, but the magnetization curve indicates that the spins are not fully polarized yet. 
Thus, we expect that the low-temperature specific heat above 0.03 T arises from the gapless 1D ferromagnetic dispersion.  
As shown in the inset of Fig.4(a), in the low-temperature region, the magnetic specific heats at 0.03 T and 0.1 T actually show $\sqrt{T}$ behavior, which clearly demonstrates the contribution of $k^2$ dispersion in 1D ferromagnetic chain~\cite{BF,ferro2,ferro3,ferro4}.

\section{Summary}
We successfully synthesized single crystals of the verdazyl-based complex $[$Zn(hfac)$_2]$$[$4-Cl-$o$-Py-V-(4-F)$_2]$. 
$Ab$ $initio$ MO calculations indicated the formation of an $S$=1/2 ferromagnetic chain. 
The magnetic susceptibility and specific heat indicated a phase transition to an AF order owing to weak interchain interactions, and the low-temperature magnetization curve exhibited a linear increase.
We explained the magnetic susceptibility above the phase transition temperature on the basis of the $S$=1/2 ferromagnetic Heisenberg chain with an AF mean field by combining the QMC method and mean-field approximation.
The magnetization curve in the ordered phase was described by the conventional AF two-sublattice model considering the AF interchain couplings.
Furthermore, we reproduced the almost temperature-independent magnetic specific heat for the $S$=1/2 ferromagnetic Heisenberg chain.
In the low-temperature region, we observed $\sqrt{T}$ dependence of the magnetic specific heat, which indicates the contribution of $k^2$ dispersion in the 1D ferromagnetic chain.

\begin{acknowledgments}
This research was partly supported by Grant for Basic Science Research Projects from KAKENHI (No. 17H04850 and No. 18H01164) and the Matsuda Foundation.
A part of this work was performed as the joint-research program of ISSP, the University of Tokyo.
\end{acknowledgments}



\begin{thebibliography}{99}

\bibitem{1DAF}
T. Giamarchi, Quantum Physics in One Dimension, International
Series of Monographs on Physics (Oxford Science, Clarendon,
Oxford, 2004).

\bibitem{bete} 
H. Bethe, Z. Phys. \textbf{71}, 205, (1931).

\bibitem{BF} 
J. C. Bonner and M. E. Fisher, Phys. Rev. \textbf{135}, A640, (1964).

\bibitem{takahashi} 
M. Takahashi, Phys. Rev. Lett. \textbf{58}, 168, (1987).

\bibitem{ferro1} 
M. Takahashi and M. Yamada, J. Phys. Soc. Jpn. \textbf{54}, 2808, (1985).

\bibitem{ferro2} 
M. Yamada and M. Takahashi, J. Phys. Soc. Jpn. \textbf{55}, 2024, (1986).

\bibitem{ferro3} 
P. Schlottmann, Phys. Rev. Lett. \textbf{54}, 2131, (1985).

\bibitem{ferro4} 
M. Shiroishi and M. Takahashi, Phys. Rev. Lett. \textbf{89}, 117201, (2002).

\bibitem{ferro_BEC} 
A. V. Syromyatnikov, Phys. Rev. B \textbf{75}, 134421, (2007).


\bibitem{TMSO_PRB} 
D. D. Swank, C. P. Landee, and R. D. Willett, Phys. Rev. B \textbf{20}, 2154, (1979).

\bibitem{DMSO_JAP} 
C. P. Landee and R. D. Willett, J. Appl. Phys. \textbf{52}, 2240, (1981).

\bibitem{CHAB_PRB} 
K. Kopinga, A. M. C. Tinus, and W. J. M. de Jonge, Phys. Rev. B \textbf{25}, 4685, (1982).

\bibitem{DMSO_neut} 
S. K. Satija, J. D. Axe, R. Gaura, R. Willett, and C. P. Landee, Phys. Rev. B \textbf{25}, 6855, (1982).

\bibitem{CHAB_neut} 
G. C. de Vries, E. Frikkee, K. Kakurai, M. Steiner, B. Dorner, K. Kopinga, and W. J. M. de Jonge, Physica B \textbf{156}, 266, (1989).

\bibitem{b-p-NPNN}
M. Tamura, Y. Nakazawa, D. Shiomi, K. Nozawa, Y. Hosokoshi, M. Ishikawa, M. Takahashi, and M. Kinoshita, Chem. Phys. Lett. \textbf{186}, 401, (1991).

\bibitem{r-p-NPNN}
M. Takahashi, P. Turek, Y. Nakazawa, M. Tamura, K. Nozawa, D. Shiomi, M. Ishikawa, and M. Kinoshita, Phys. Rev. Lett. \textbf{67}, 746, (1991).

\bibitem{b-r-both}
Y. Nakazawa, M. Tamura, N. Shirakawa, D. Shiomi, M. Takahashi, M. Kinoshita, and M. Ishikawa, Phs. Rev. B \textbf{46}, 8906, (1992).

\bibitem{BImNN_CL}
N. Yoshioka, M. Irisawa, Y. Mochizuki, T. Kato, H. Inoue, and S. Ohba, Chem. Lett. \textbf{26}, 251, (1997).

\bibitem{BImNN_uSR}
T. Sugano, S. J. Blundell, T. Lancaster, F. L. Pratt, and H. Mori, Phys. Rev. B \textbf{82}, 180401(R), (2010).


\bibitem{F4BImNN_JACS}
H. Murata, Y. Miyazaki, A. Inaba, A. Paduan-Filho, V. Bindilatti, N. F. Oliveira Jr., Z. Delen, and P. M. Lahti, J. Am. Chem. Soc. \textbf{130}, 186, (2008).

\bibitem{F4BImNN_uSR}
S. J. Blundell, J. S. M$\ddot{\rm{o}}$ller, T. Lancaster, P. J. Baker, F. L. Pratt, G. Seber, and P. M. Lahti, Phys. Rev. B \textbf{88}, 064423, (2013).

\bibitem{BBDTA_PRB}
K. Shimizu, T. Gotohda, T. Matsushita, N. Wada, W. Fujita, K. Awaga, Y. Saiga, and D. S. Hirashima, Phys. Rev. B \textbf{74}, 172413, (2006).



\bibitem{a26Cl2V} 
H. Yamaguchi, T. Okubo, S. Kittaka, T. Sakakibara, K. Araki, K. Iwase, N. Amaya, T. Ono, and Y. Hosokoshi, Sci. Rep. {\bf 5}, 15327 (2015).


\bibitem{random} 
H. Yamaguchi, M. Okada, Y. Kono, S. Kittaka, T. Sakakibara, T. Okabe, Y. Iwasaki, Y. Hosokoshi, Sci. Rep. {\bf 7}, 16144 (2017).

\bibitem{TCNQ_square} 
H. Yamaguchi, Y. Tamekuni, Y. Iwasaki, Y. Hosokoshi, Phys. Rev. B {\bf 97}, 201109(R) (2018).

\bibitem{3Cl4FV} 
H. Yamaguchi, K. Iwase, T. Ono, T. Shimokawa, H. Nakano, Y. Shimura, N. Kase, S. Kittaka, T. Sakakibara, T. Kawakami, and Y. Hosokoshi, Phys. Rev. Lett. {\bf 110}, 157205 (2013).




\bibitem{2Cl6FV} 
H. Yamaguchi, A. Toho, K. Iwase, T. Ono, T. Kawakami, T. Shimokawa, A. Matsuo, and Y. Hosokoshi, J. Phys. Soc. Jpn. {\bf 82}, 043713 (2013).


\bibitem{PF6} 
H. Yamaguchi, Y. Sasaki, T. Okubo, M. Yoshida, T. Kida, M. Hagiwara, Y. Kono, S. Kittaka, T. Sakakibara, M. Takigawa, Y. Iwasaki, and Y. Hosokoshi, Phys. Rev. B {\bf 98}, 094402 (2018).

\bibitem{3IV_kono} 
Y. Kono, S. Kittaka, H. Yamaguchi, Y. Hosokoshi, and T. Sakakibara, Phys. Rev. B {\bf 97}, 100406(R) (2018).
  
\bibitem{Kuhn}
R. Kuhn and H. Trischmann, Monatsh. Chem. {\bf 95}, 457 (1964).

\bibitem{Xray}
G. M. Sheldrick, Acta Crystallogr., Sect. C: Struct. Chem. {\bf C71}, 3 (2015).


\bibitem{MOcal} M. Shoji, K. Koizumi, Y. Kitagawa, T. Kawakami, S. Yamanaka, M. Okumura, and K. Yamaguchi, Chem. Phys. Lett. {\bf 432}, 343 (2006).


\bibitem{QMC2}  A. W. Sandvik, Phys. Rev. B {\bf 59}, 14157 (1999).

\bibitem{ALPS}  A. F. Albuquerque, F. Alet, P. Corboz, P. Dayal, A. Feiguin, L. Gamper, E. Gull, S. Gurtler, A. Honecker, R. Igarashi, M. Korner, A. Kozhevnikov, A. Lauchli, S. R. Manmana, M. Matsumoto, I. P. McCulloch, F. Michel, R. M. Noack, G. Pawlowski, L. Pollet, T. Pruschke, U. Schollwock, S. Todo, S. Trebst, M. Troyer, P. Werner, and S. Wessel, J. Magn. Magn. Mater. {\bf 310}, 1187 (2007) (see also http://alps.comp-phys.org and http://wistaria.comp-phys.org/alps-looper/).


\bibitem{ALPS3}B. Bauer, L. D. Carr, A. Feiguin, J. Freire, S. Fuchs, L. Gamper, J. Gukelberger, E. Gull, S. Guertler, A. Hehn, R. Igarashi, S.V. Isakov, D. Koop, P.N. Ma, P. Mates, H. Matsuo, O. Parcollet, G. Pawlowski, J.D. Picon, L. Pollet, E. Santos, V.W. Scarola, U. Schollw$\ddot{\rm{o}}$ck, C. Silva, B. Surer, S. Todo, S. Trebst, M. Troyer, M.L. Wall, P. Werner, and S. Wessel, J. Stat. Mech.: Theory and Experiment, P05001 (2011).


\bibitem{MOseido} T. Kawakami, Y. Kitagawa, F. Matsuoka, Y. Yamashita,
and K. Yamaguchi, Polyhedron {\bf 20}, 1235 (2001).












\end{thebibliography}
\end{document}